\documentclass[
 reprint,
 superscriptaddress, longbibliography,
 amsmath,amssymb,
]{revtex4-2}

\usepackage{gensymb}
\usepackage{float}
\usepackage{siunitx}
\usepackage{lipsum}
\usepackage{amsmath}
\usepackage{graphicx}
\usepackage[colorlinks=true, allcolors=blue]{hyperref}
\usepackage{dcolumn}
\usepackage{bm}
\usepackage{color}

\begin{document}

\title{Simultaneous imaging of dopants and free charge carriers by STEM-EELS}

\author{Hongbin Yang}
\email{hongbin.yang@rutgers.edu}
\affiliation{Department of Chemistry and Chemical Biology, Rutgers University, Piscataway, New Jersey, 08854, USA}

\author{Andrea Kone\v{c}n\'{a}}
\affiliation{ICFO-Institut de Ciencies Fotoniques, The Barcelona Institute of Science and Technology, 08860 Castelldefels, Barcelona, Spain}
\affiliation{Central European Institute of Technology, Brno University of Technology, 61200 Brno, Czech Republic}

\author{Xianghan Xu}
\author{Sang-Wook Cheong}
\affiliation{Department of Physics and Astronomy, Rutgers University, Piscataway, New Jersey, 08854, USA}

\author{Philip E. Batson}
\email{batson@physics.rutgers.edu}
\affiliation{Department of Physics and Astronomy, Rutgers University, Piscataway, New Jersey, 08854, USA}

\author{F. Javier {Garc\'{i}a de Abajo}}
\email{javier.garciadeabajo@nanophotonics.es}
\affiliation{ICFO-Institut de Ciencies Fotoniques, The Barcelona Institute of Science and Technology, 08860 Castelldefels, Barcelona, Spain}
\affiliation{ICREA-Instituci\'{o} Catalana de Recerca i Estudis Avan\c{c}ats, Passeig Llu\'{i}s Companys 23, 08010 Barcelona, Spain}

\author{Eric Garfunkel}
\affiliation{Department of Chemistry and Chemical Biology, Rutgers University, Piscataway, New Jersey, 08854, USA}
\affiliation{Department of Physics and Astronomy, Rutgers University, Piscataway, New Jersey, 08854, USA}
\date{\today}

\begin{abstract}
Doping inhomogeneities in solids are not uncommon, but their microscopic observation and understanding are limited due to the lack of bulk-sensitive experimental techniques with high-enough spatial and spectral resolution. Here, we demonstrate nanoscale imaging of both dopants and free charge carriers in La-doped $\mathrm{BaSnO_3}$ (BLSO) using high-resolution electron energy-loss spectroscopy (EELS). By analyzing both high- and low-energy excitations in EELS, we reveal chemical and electronic inhomogeneities within a single BLSO nanocrystal. The inhomogeneous doping leads to distinctive localized infrared surface plasmons, including a novel plasmon mode that is highly confined between high- and low-doping regions. We further quantify the carrier density, effective mass, and dopant activation percentage from EELS data and transport measurements on the bulk single crystals of BLSO. These results represent a unique way of studying heterogeneities in solids, understanding structure-property relationships at the nanoscale, and opening the way to leveraging nanoscale doping texture in the design of nanophotonic devices.
\end{abstract}

\maketitle
\section{Introduction}

Electrical and optical properties of materials are strongly influenced by the density of free charge carriers~\cite{Zunger2021ChemRev}. Consequently, tuning the carrier density has been of critical importance for the design and fabrication of semiconductor devices, as well as for exploring novel electronic states in complex oxides such as superconductivity~\cite{Ye2012Science,Edge2015PRL,Li2020PRL} and topological effects~\cite{Hong2012NatureComm}. Chemical doping is often exploited to tune the carrier density in solids, although it may introduce chemical and electronic inhomogeneities at the nanoscale~\cite{Perea2009NatureNanotechnology,Koren2010Nanoletter,Ko2022AM}. These inhomogeneities are often detrimental to carrier mobility and device performance~\cite{Weber2012Science, Jariwala20182DMater, Jagadamma2021ACSAEM}. While in some material systems, the inevitable heterogeneities exhibit close correlation with the emergence of ferroelectricity~\cite{Manley2014NatureComm, Li2016NatureComm, Kumar2021NatureMaterials}, high-temperature superconductivity~\cite{Pan2001Nature,Hanaguri2004Nature,Vershinin2004Science}, and unconventional magnetism~\cite{Uehara1999Nature, Burgy2001PRL, Akahoshi2003PRL, Dagotto2005Science}.

The relationship between these exotic properties and nanoscale heterogeneities has been challenging to study because of the high demand on both spatial and spectral resolution of the required experimental techniques. 
Scanning probe microscopy is potentially useful to cover this gap because it offers the possibility to investigate a wide range of local phenomena, including surface structures and local density of electronic states by scanning tunneling microscopy~\cite{Yin2021NatureRevPhys}, nanoscale electrical properties by electrostatic force microscopy~\cite{Girard2001Nanotech, Gramse2020NatureElectronics}, and optical excitation strengths by scanning near-field optical microscopy~\cite{Johns2016NatureComm,Mooshammer2018NanoLetter,Lu2018AEM,Chen2019AM, Chu2020AnnRevMaterRes}. However, these scanning probe techniques are primarily sensitive to surface properties, and only offer access to a limited spectral range. Thus, it remains difficult to directly relate the local structure and chemical composition to the electrical or optical properties in materials of interest~\cite{Weber2012Science, Cojocaru2017SciRep}. In this regard, scanning transmission electron microscopy (STEM) combined with electron energy-loss spectroscopy (EELS) represent a powerful alternative because they allow imaging of atomic structures with simultaneous measurement of local elemental composition, chemical bonding, and electronic excitations down to the single-atom level~\cite{ZLN12}.

Thanks to a series of recent advances in energy resolution~\cite{Krivanek2014Nature, Dellby2020MM}, state-of-the-art STEM-EELS provides access into excitations over a broad energy range (from a few $\mathrm{meV}$ to $\,\mathrm{keV}$'s \cite{Krivanek2019Ultramicroscopy}), therefore emerging as the technique of choice to simultaneously probe high-energy core-level electronic transitions and low-energy excitations, such as phonons and vibrational signals~\cite{Lagos2017Nature, Govyadinov2017NatureCommunications, Hage2018SciAdv, Senga2019Nature, Venkatraman2019NaturePhysics, HRK20, Yan2021Nature, 2021SmallKonecna, Li2022PNAS}, infrared plasmons~\cite{Granerod2018PRB,Yang2020PRB, Olafsson2020NanoLetters, Mkhitaryan2021NanoLetters}, and valence electronic excitations~\cite{Abajo2010RMP, Tizei2015PRL}. Importantly, EELS can be combined with high-resolution imaging~\cite{Voyles2002Nature,Varela2004PRL} to reveal local structure-property relationships.
%including dramatic examples in which a single atom modifies the optical response \cite{ZLN12,HRK20}.

\begin{figure*}
\centering
\includegraphics[width=16.cm]{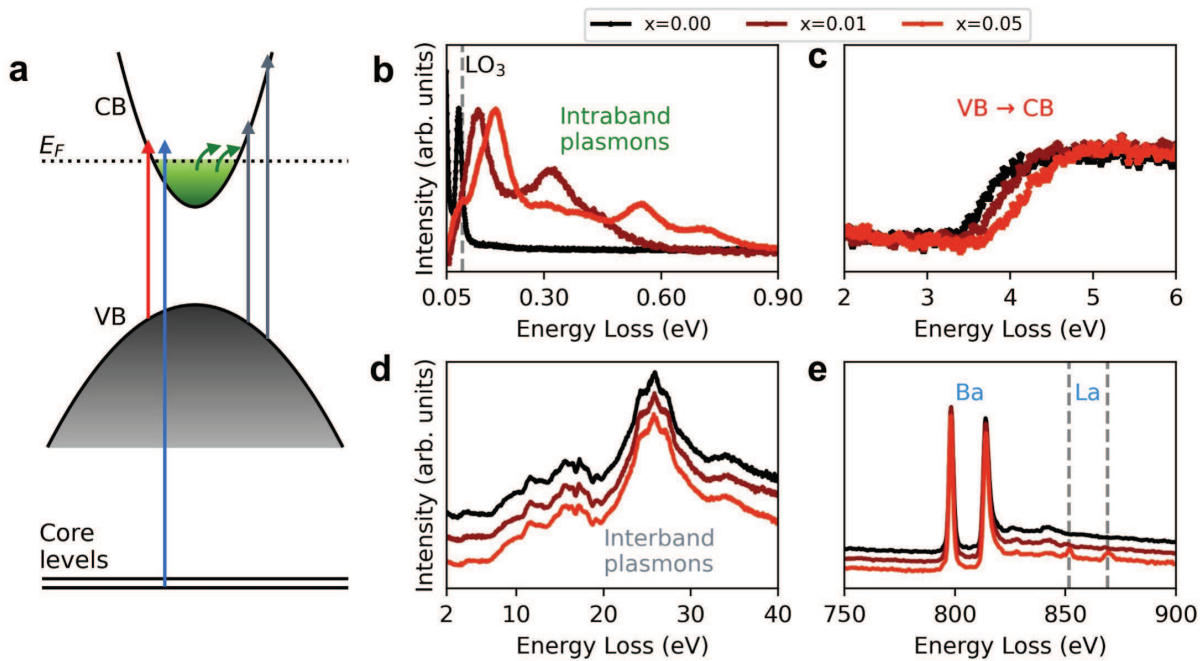}
\caption{\textbf{Electronic excitations in EELS.} \textbf{a}, Schematic diagram of electronic excitations, including collective modes (intraband plasmons of free carriers indicated by the green arrows, and interband plasmons indicated by the gray arrows) and single-particle transitions from core levels (blue) or the valence band (VB, red) to the conduction band (CB) state above $\mathrm{E_F}$. \textbf{b}-\textbf{d}, Doping dependence of low-loss and core-loss EELS in $\mathrm{Ba_{1-x}La_xSnO_3}$ single crystals with $\mathrm{x=}$ 0, 0.01, and 0.05. In \textbf{b}, we identify electron energy losses associated with the excitation of phonons and free-carrier plasmons. The vertical dashed line indicates the highest longitudinal optical phonon energy near 100\,meV. Valence-to-conduction band-edge transitions are shown in \textbf{c}. Valence plasmons and higher-energy interband transitions are studied in \textbf{d}. \textbf{e}, High-energy EELS data showing transitions from Ba and La core-levels to unoccupied states.}
\label{fig:1}
\end{figure*}

Here, we demonstrate a novel way of characterizing both chemical and electronic inhomogeneities at the nanoscale, achieved by simultaneous imaging of dopants and free charge carriers in La-doped $\mathrm{BaSnO_3}$ (BLSO), a high mobility perovskite oxide~\cite{Kim2012APE,Raghavan2016APLMaterials,Paik2017APLMaterials}. Through monochromated STEM-EELS, we measure electronic excitations associated with free carriers and dopants in the low-loss and core-loss regions of EELS, respectively. These measurements allow for direct observation of chemical and electronic inhomogeneities, further analysis also reveals the local doping characteristics even within a single nanocrystal of BLSO, thus signify the advantage of correlating low- and high-energy inelastic scatterings in EELS. Our results unambiguously show the strong influence of inhomogeneous doping on the plasmonic response. Specifically, we experimentally observe and numerically corroborate a novel interfacial plasmon mode localized between high- and low-doping regions. The present analysis also reveals inhomogeneous doping as a novel tool for plasmon customization on the few-nanometer scale, with potential application in the design of advanced devices necessitating a high surface-and-bulk-hybrid confinement of optical fields.

\section{Results}
\subsection{Doping dependence of low-loss and core-loss EELS}

In EELS, energy loss events can occur due to different mechanisms, including individual and collective electronic excitations that span a wide energy range, as schematically shown in Fig. \ref{fig:1}\textbf{a}. The spatial distribution of these excitations can be revealed thanks to the sub-nanometer size of the electron beam traversing the specimen. At relatively high energies, valence plasmons (i.e., collective electronic modes analogous to interband plasmons, but occurring at energies largely exceeding the electronic band gap) and core-losses (single-particle interband transitions involving core electrons to unoccupied states above $\mathrm{E_F}$) are routinely observed in EELS experiments, even at relatively low energy resolution~\cite{Abajo2010RMP}. Lower-energy excitations in solids, such as phonons, plasmons, as well as interband transitions in the UV-visible spectral range, can be better resolved with the aid of an electron monochromator capable of providing few-meV energy resolution~\cite{Krivanek2014Nature, Abajo2021acsPhoto}.

Here, we observe all these types of excitations in the bulk single crystals of $\mathrm{Ba_{1-x}La_xSnO_3}$ as a function of nominal doping, x, as shown in Fig. \ref{fig:1}\textbf{b}-\textbf{e}. These bulk single crystals were grown by floating zone method with homogeneous doping (Mehods). Starting from a wide band gap insulator without doping (x=0), $\mathrm{Ba_{1-x}La_xSnO_3}$ becomes a degenerate semiconductor as the level of La-doping increases (see x=0.01, 0.05 curves in Fig. \ref{fig:1}\textbf{b},\textbf{c}). The metallic transport properties are manifested in the infrared dielectric response, as revealed in the low-loss region of monochromated EELS in Fig. \ref{fig:1}\textbf{b}. In the 50-900\,meV spectral window, the main processes contributing to electron energy losses are lattice vibrations (phonons) and plasmons associated with the collective oscillations of doping charge carriers. For undoped crystals, we observe only narrow resonances below $\sim$90\,meV, corresponding to the phonon polaritons in $\mathrm{BaSnO_3}$~\cite{Stanislavchuk2012JAP}. With increasing n-type doping, the charge carriers have higher Fermi velocities and respond faster to the external field than the ions. The Coulomb interaction between fast electrons and free carriers thus give rise to the plasmons that we observe in doped $\mathrm{BaSnO_3}$~\cite{Pines1952PR}. We refer to these low-energy excitations as carrier plasmons in what follows. As elaborated in later sections, the spectral range of carrier plasmons is primarily set by the carrier density. For the doping level studied here, we observe surface and bulk carrier plasmons below about 1\,eV~\cite{Yang2022Small}. 

Doping-induced changes in the optical band gap, the Burstein-Moss (B-M) shift~\cite{Burstein1954PhysicalReview, Moss1954PPSB}, is also prominent in our EELS measurements over the 2-6\,eV energy range in Fig. \ref{fig:1}\textbf{c}. $\mathrm{BaSnO_3}$ has a wide band gap $>3\,$eV~\cite{Monserrat2018PRB} and a sufficiently small high-frequency dielectric constant $\epsilon_\infty$ for it to make Cherenkov losses negligible in our EELS experiments with a 60\,keV electron beam~\cite{Chen1975PRB, Stoger-Pollach2006Micron}. Although $\mathrm{BaSnO_3}$ has a smaller indirect band gap, the associated transition probability is orders of magnitude lower than the direct transitions~\cite{Monserrat2018PRB}. Thus, the energy loss intensities we observe here primarily stem from direct interband transitions near the $\Gamma$ points~\cite{Aggoune2022CommMaterials}. The loss signal from indirect transitions likely overlaps with that from defect states from the top and bottom surfaces of the TEM specimen~\cite{Rafferty1998PRB}, thus appearing as a non-zero background below the absorption onset dictated by the direct band-gap energy. With increasing doping, a blue shift of the optical gap is clearly observed in Fig. \ref{fig:1}\textbf{c}.

\begin{figure*}
\centering
\includegraphics[width=16.cm]{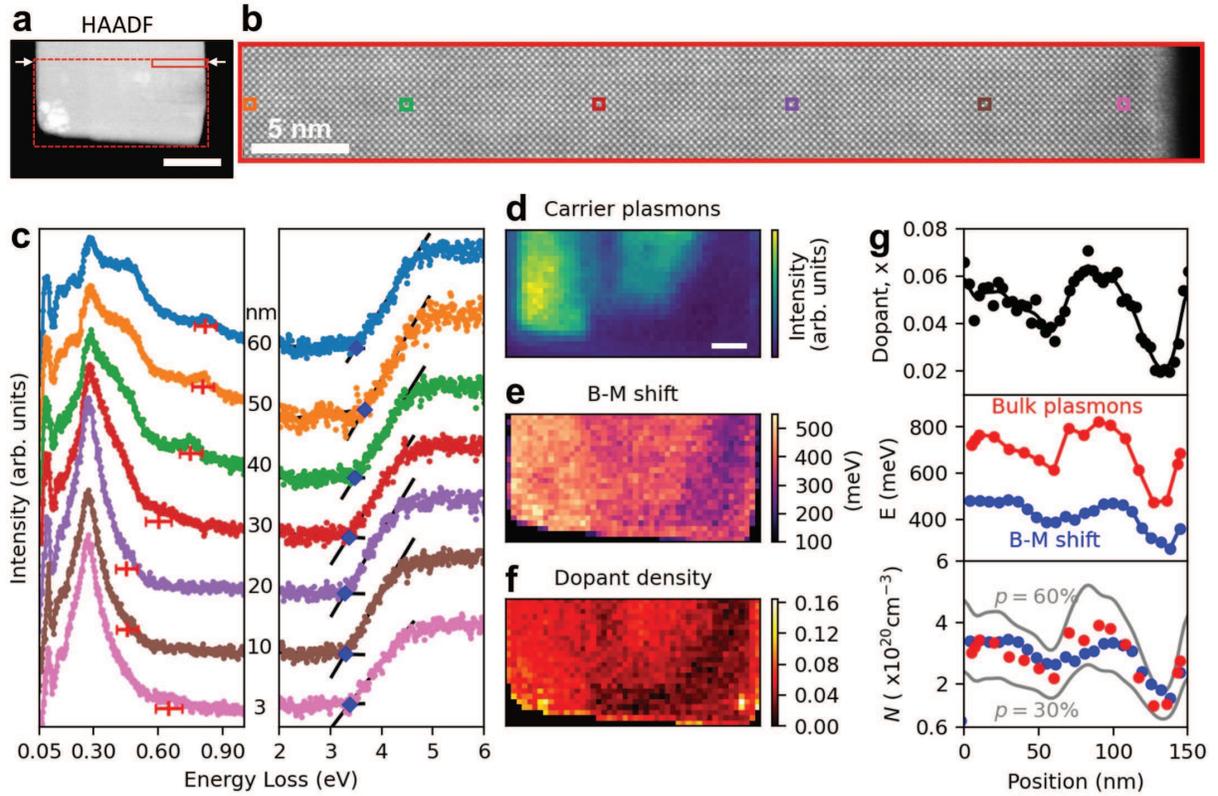}
\caption{\textbf{Mapping chemical and electronic inhomogeneities in a BLSO nanocrystal.} \textbf{a}, STEM-HAADF image of a cubic-shaped BLSO nanocrystal. \textbf{b}, Atomic-resolution HAADF image near the edge of the crystal, within the rectangular region marked by solid red lines in \textbf{a}. \textbf{c}, Low-loss EELS signal acquired at electron beam positions indicated by color-matching squares in the HAADF image in \textbf{b}. The corresponding distances to the crystal edge are labeled between the two sets of spectra. Spectra at consecutive distances are offset for clarity. The spectrum at 60\,nm (blue) is taken outside the region shown in \textbf{b}. Red and blue markers indicate energies of bulk carrier plasmons and band gaps, respectively. \textbf{d}, Energy-filtered maps of bulk plasmons integrated from 780 to 830\,meV. \textbf{e}, Burstein-Moss shift map, and \textbf{f}, dopant percentage map for the nanocrystal in \textbf{a} over the spatial range indicated by the dashed rectangular boundary. \textbf{g}, Dopant percentage (top), bulk carrier plasmon energy and B-M shift (middle), and carrier density (bottom) as a function of electron beam position along a horizontal line between the two white arrows in \textbf{a}. Gray curves in bottom panel of \textbf{g} indicate doping activation percentages of 30\% and 60\%. The scale bars in \textbf{a} and \textbf{d} indicate 50\,nm and 20\,nm, respectively.
}
\label{fig:2}
\end{figure*}

It should be noted that the carrier plasmons with a strong doping-dependence in Fig. \ref{fig:1}\textbf{b} are intraband in nature. They are accompanied by other more lossy interband plasmons at relatively-high energies, whose energy is in fact proportional to the square root of the valence electron density. In view of their origin, we denote these modes as valence plasmons, in contrast to the aforementioned sub-eV carrier plasmons. In the case of undoped $\mathrm{BaSnO_3}$, valence plasmon is around $\sim26\,$eV~\cite{Yun2018JVSTA}. In our results, valence plasmons and high-energy interband transitions both show no measurable dependence on doping--we observe nearly identical spectra in undoped and doped crystals, see Fig. \ref{fig:1}\textbf{d}. This is not unexpected because the lattice constant and valence electronic band structure remain largely unaffected when substituting La for Ba. The introduced carrier density is typically below \SI{1E21}{cm^{-3}}~\cite{Prakash2017NatureComm}, which is orders of magnitude smaller than the valence electron density and should accordingly cause only marginal variations in the valence plasmons at about 26\,eV. Therefore, changes due to carrier doping can be measured exclusively in the low-loss EELS region shown in Fig. \ref{fig:1}\textbf{b} and \textbf{c}.

We also analyzed core-loss EELS associated with the A-site chemical composition of $\mathrm{BaSnO_3}$ (Fig. \ref{fig:1}\textbf{e}). As expected, the La-$\mathrm{M_{4,5}}$ edge intensity increases with nominal doping. For this material system and many others, where the dopant and host lattice differ in atomic number by only one, core-loss EELS is more useful than high-angle annular dark field (HAADF) imaging for identifying dopant locations. 

\subsection{Real-space mapping of dopants and charge carriers}

From the knowledge of the energies of high- and low-energy excitations shown above, dopants and free carriers can be visualized in real space by collecting energy-filtered EELS maps. We demonstrate such mapping for a single cubic BLSO nanocrystal, as shown in Fig. \ref{fig:2}\textbf{a}. Although atomic-resolution HAADF imaging of the crystal in Fig. \ref{fig:2}\textbf{b} shows no substantial contrast variation, low-loss EEL spectra in Fig. \ref{fig:2}\textbf{c} reveal significant changes due to electronic inhomogeneities. From the spectra in the 50-1000\,meV energy range, we see how the bulk carrier plasmons are shifting from 830\,meV (at a distance of 50\,nm from the flake edge) to about 500\,meV (20\,nm away from the edge). Bulk plasmon energies are signaled by the zero-crossing of the frequency-dependent BLSO dielectric function $\epsilon(\omega)$ (i.e., a pole in the loss function $\mathrm{Im}\{-1/\epsilon(\omega)\}$). These modes show up as isolated peak maxima in the loss spectra for electron-beam placed at 60, 50, 40, 3\,nm from the flake edge, while they emerge as a shoulder of the more intense surface plasmons in the spectra acquired at distances of 30, 20, 10\,nm. This is because with decreasing energy, the bulk plasmon peaks not only redshift in energy, but also decrease in intensity~\cite{Caruso2018PRB} (Section~S1, 2, Supporting Information). In the 200-600\,meV spectral range, we observe surface plasmons whose energies are affected by both the local plasma frequency and the specimen shape, as we show in Fig. \ref{fig:3} below. In addition, we observe phonon polaritons between 80-100\,meV, whose intensities are inversely proportional to the bulk plasmon energy (Section~S1, Supporting Information). These phonon polaritons stem from regions where lattice vibrations are not completely screened by free carriers, such as in surface depletion layers~\cite{Chen1989PRB, Chen1989PRB2}. 

Local optical gaps measured by EELS also show spatial variations, as shown the in 2-6\,eV spectra in Fig. \ref{fig:2}\textbf{c}. We extract the optical gap by modeling the absorption onset (Section~S2, Supporting Information), and find that the local optical gap changes between 3.28\,eV and 3.66\,eV within the region shown in Fig. \ref{fig:2}\textbf{b}, depending on the electron beam positions.

Carrier plasmons and B-M shifts are visualized in the EELS maps in Fig. \ref{fig:2}\textbf{d} and \ref{fig:2}\textbf{e}, from which we find clear inhomogeneities across the whole nanocrystal. Namely, the signal associated with bulk carrier plasmons (integrated over the 780-830\,meV) is observed only in part of the flake (left and center-top). The B-M shift (evaluated with respect to the value obtained for x=0) also shows significant spatial variation, in the range of 200-550\,meV. The maps of these two quantities exhibit close correlations, suggesting that their variations are both related with changes in the doping level of the free carriers. We also reveal that these electronic inhomogeneities are closely related to the dopant percentage in Fig. \ref{fig:2}\textbf{f}, acquired from core-loss EELS maps. Here, the dopant percentage x is quantified via the the k-factor method, based on the ratio of Ba- and La-$\mathrm{M_{4,5}}$-edge cross sections~\cite{Hofer1988Ultramicroscopy}. We see that the areas with larger x coincide with regions exhibiting a large B-M shift. These regions also feature the high-energy bulk carrier plasmons. In addition, the dopant percentage map in  Fig. \ref{fig:2}\textbf{f}, although obtained by analyzing a different energy range, shows close correlation between the spatial distribution of chemical and electronic inhomogeneities. These spatial variations are also compared more quantitatively in Fig.~\ref{fig:2}\textbf{g}, where we plot line profiles of the dopant percentage, the bulk plasmon energy, and the B-M shift as a function of beam position.

\subsection{Influence of doping inhomogeneity on localized surface plasmons}

\begin{figure*}
\centering
\includegraphics[width=18cm]{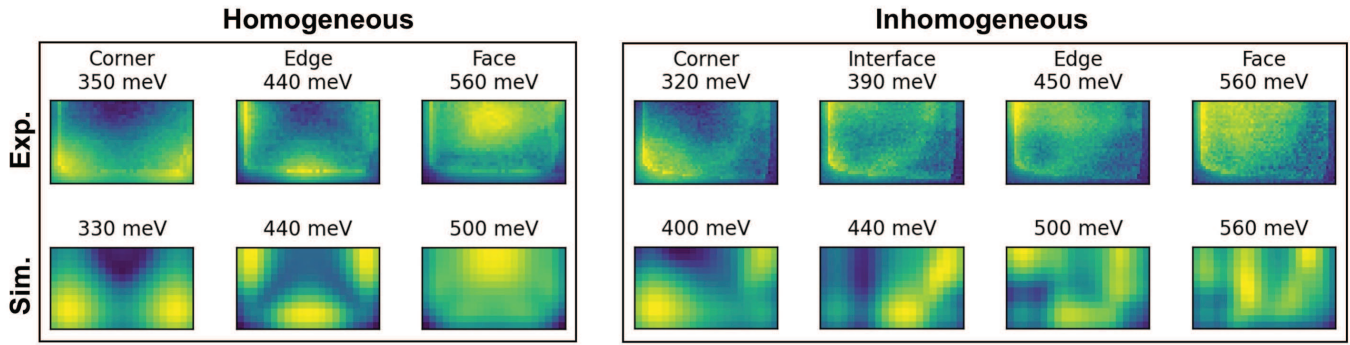}
\caption{\textbf{Localized surface plasmons in BLSO nanocrystals.} Experimental (top) and simulated (bottom) localized surface plasmons in cubic BLSO flakes with homogeneous (left) and inhomogeneous (right) carrier density distributions.}
\label{fig:3}
\end{figure*}

The doping inhomogeneity determined above has a strong influence on the global plasmonic response of the BLSO nanocrystal. This can be seen in a comparison between BLSO nanocrystals with homogeneous and inhomogeneous carrier density distribution, as illustrated in Fig. \ref{fig:3}. In the left panel of Fig. \ref{fig:3}, a nearly cubic BLSO nanoparticle with a relatively homogeneous carrier density distribution shows localized surface plasmon resonances near the corners, edges, and faces at around 350\,meV, 450\,meV, and 560\,meV, respectively. The energy and spatial distribution of these surface plasmons are well reproduced by numerical calculations. Similar geometrically-controlled modes are characteristic in cubic-shaped particles, as previously observed in both plasmonic~\cite{Scherry2005NanoLetters,Nicoletti2013Nature,Zhang2011NanoLetters} and phononic~\cite{Lagos2017Nature,Li2021Science} materials.

In contrast to homogeneous particles, the BLSO nanocrystals under study, featuring inhomogeneous doping, exhibit substantially different surface plasmon resonances. In the right panel of Fig. \ref{fig:3}, we visualize the localized surface plasmons of the BLSO nanoparticle studied in Fig. \ref{fig:2}. Instead of having corner mode plasmons that brighten all corners at one resonance energy, we observe resonances localized near the left and right corner at different energies. The corner mode at 320\,meV only appears near the left corner. The right corner also shows a localized plasmon, but at a much lower energy $\sim180\,$meV (Section~S3, Supporting Information). Although this particle has almost identical size and shape as the one in the left panel of Fig. \ref{fig:3}, the inhomogeneity in local dopant percentage leads to a different spatially varying carrier density, and thus bulk plasma frequency. We corroborate these experimental results through numerical simulations, where we use locally-varying bulk plasmon frequencies extracted from experiments. The simulated results also reveal corner modes occurring at different energies as well as similar patterns in energy-filtered maps corresponding to the edge- and face-like plasmons around 450\,meV and 560\,meV, respectively. The relatively minor differences between experimental and theoretical results, particularly visible at higher energies, can stem from additional inhomogeneities in plasmon damping, not incorporated in the numerical models.

Interestingly, a unique interfacial plasmon mode is found only in the inhomogeneous nanocrystal. As shown in Fig. \ref{fig:3} (right panel), this mode concurs with the left corner resonance near 320\,meV, which is more pronounced with increasing energy and is peaked at about 390\,meV, before finally merging with the face-like mode resonance near the high-doping side (above 450\,meV in experiments). This interfacial mode forms a circular pattern that is highly confined between the low- and high-doping regions across the whole particle. Although the dopant positions here are random, this phenomenon points to the possibility of engineering novel plasmonic properties by tuning chemical doping at the atomic scale. For instance, electromagnetic wave could be guided \cite{Law2014PRL} along an arbitrary path in a low-doping channel with extremely confined size that is determined by the placement of dopants.

\subsection{Quantifying free carrier characteristics and doping efficiency}

\begin{figure*}
\centering
\includegraphics[width=18cm]{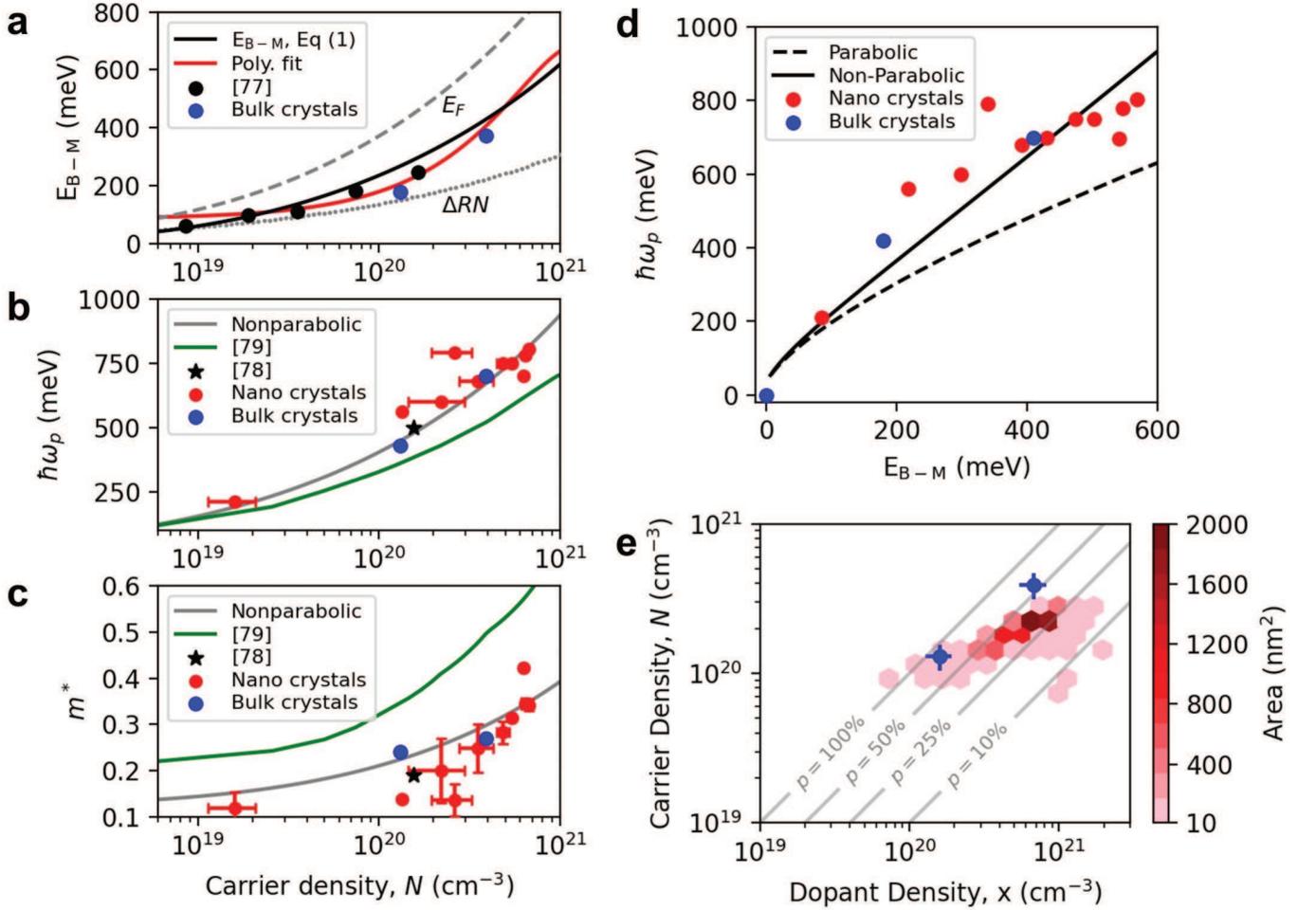}
\caption{\textbf{Free carrier characteristics and doping efficiency.} Dependence of \textbf{a}, Burstein-Moss (B-M) shift, \textbf{b}, bulk plasma frequency, and \textbf{c}, carrier effective mass on carrier density. Results from Refs.~\cite{Lebens-Higgins2016PRL, Allen2016APL, Rowberg2020PRB} are shown in comparison. \textbf{d}, Bulk plasmon energy as a function of B-M shift for bulk and nano crystals as extracted from experiments (symbols), compared with modeled results (black curves). \textbf{e}, Carrier concentration as a function of dopant percentage in BLSO. The carrier densities of bulk single crystals (blue dots) are obtained from Hall measurements at room temperature. Carrier and dopant densities of the nanocrystals are extracted from EELS maps in Fig. \ref{fig:2}. Data points from each 3.2 $\times$ 3.2 $\mathrm{nm^2}$ pixel in the EELS maps are represented by hexagons, colored according to the total area for a given $N$ and x (see color scale).}
\label{fig:4}
\end{figure*}

Next, we quantify the carrier density from the experimentally measured bulk plasma frequency and optical band gap without making any assumption on the carrier effective mass. The bulk plasma frequency takes the form $\omega_{\rm p} = \sqrt{{Ne^2}/{\epsilon_0 m^*}}$, where $N$ is the carrier density and $m^*$ the effective mass. Although the bulk plasma frequency is informative about the infrared and optical properties of materials, it only reveals the ratio between carrier density and effective mass~\cite{Franzen2008JPCC}. Carrier effective mass is often highly dependent on doping level because of band non-parabolicity. Thus, in order to unambiguously determine the carrier density and effective mass from a measurement of $\hbar\omega_{\rm p}$, an additional doping-dependent quantity needs to be measured as well. In this work we find that the simultaneous measurement of bulk plasmon loss and the Burstein-Moss shift by EELS makes it possible to quantify both the local carrier density and the effective mass.

For the sake of corroboration, to quantify the free carrier density and effective mass from EELS, we have also performed Hall measurements on the bulk single crystals without significant inhomogeneities, whose bulk plasmon energy and B-M shift were studied by EELS in Fig.~\ref{fig:1}. The BLSO bulk crystals with $\mathrm{x=}$ 0.01 and 0.05 have carrier densities $N$ of \SI{1.3e20}{cm^{-3}} and \SI{3.9e20}{cm^{-3}}, respectively. The results are shown in Fig.~\ref{fig:4}\textbf{a}, together with data from a previous study of BLSO thin films by optical and Hall experiments~\cite{Lebens-Higgins2016PRL}. We see that the B-M shift increases with carrier density, following a similar trend, although the data are measured with different techniques. The dependence of the B-M shift on carrier density is determined directly from the electronic band structure of $\mathrm{BaSnO_3}$. Specifically, the B-M shift can be modeled based on the conduction band structure and doping level as~\cite{Berggren1981PRB}
\begin{equation}
    E_\mathrm{B-M}= E_\mathrm{F}-\Delta_\mathrm{RN}, \nonumber
\end{equation}
where $E_\mathrm{F}$ is the Fermi energy relative to the conduction band minimum and $\Delta_\mathrm{RN}$ is the gap renormalization due to electron-electron and electron-impurity interactions. More precisely, $\Delta_\mathrm{RN}= \Delta E_\mathrm{g}^\mathrm{ee} + \Delta E_\mathrm{g}^\mathrm{ei}$, where
\begin{equation}
    \Delta E_\mathrm{g}^\mathrm{ee} = \frac{e^2k_\mathrm{F}}{2\pi^2\epsilon_s\epsilon_0}+
    \frac{e^2k_\mathrm{TF}}{8\pi\epsilon_s\epsilon_0}\left[1-\frac{4}{\pi}\tan^{-1}\left(\frac{k_\mathrm{F}}{k_\mathrm{TF}}\right)\right] \nonumber
\end{equation}
and
\begin{equation}
    \Delta E_\mathrm{g}^\mathrm{ei} = \frac{Ne^2}{\epsilon_s\epsilon_0a_B^*k_\mathrm{TF}^3}. \nonumber
\end{equation}
Here, $e$ is the elementary charge, $\epsilon_0$ is the vacuum permittivity, $\epsilon_\mathrm{s}$ is static relative dielectric permittivity, $a_\mathrm{B}^*=\frac{\epsilon_\mathrm{s}}{m^*}a_0$ is the effective Bohr radius, $a_0$ is the Bohr radius, $m^*$ is the effective mass, $k_\mathrm{TF}=\sqrt{\frac{4k_\mathrm{F}}{\pi a_\mathrm{B}^*}}$ is the Thomas-Fermi screening wave vector, and $k_\mathrm{F}=(3\pi N)^{1/3}$ is the Fermi wave vector. Considering a first order non-parabolic model, the effective mass follows
\begin{equation}
    m^* = m^*_{E_\mathrm{F}=0}\sqrt{1+2\beta\frac{\hbar^2k^2}{m^*_{E_\mathrm{F}=0}}}, \nonumber
\end{equation}
where $m^*_{E_\mathrm{F}=0}$ is the carrier effective mass at the conduction band bottom, and $\beta$ is a parameter describing the band curvature (the increase in $m^*$ with electron wave vector $k$). 

We model the conduction band of $\mathrm{BaSnO_3}$ by varying the carrier effective mass and, thus, the degree of non-parabolicity. Additional variables are the static and high frequency dielectric constants, $\epsilon_\mathrm{s}$ and $\epsilon_\infty$, which do not change significantly with doping. The results are compared with the experimentally measured $E_{B-M}$ and $N$ in Fig. \ref{fig:4}\textbf{a}. Alternatively, a polynomial fit of $E_\mathrm{B-M}(N)$ (red curve in Fig.~\ref{fig:4}\textbf{a}) can also provide information on the conduction band curvature. The mean value of the non-parabolic conduction band model and the polynomial fit is shown in Fig.~\ref{fig:4}\textbf{b} and \ref{fig:4}\textbf{c}. Our best fit to experimental results were found with $m^*_{E_{F}=0}$ = 0.12$\pm0.2$ and $\beta$=1.4 for the conduction band of $\mathrm{BaSnO_3}$.

Electron effective mass in $\mathrm{BaSnO_3}$ was found in a wide range in earlier studies, with the smallest value being $m^*=$ 0.06~\cite{Moreira2012JSSC} and most experimental results indicating $m^*$ close to 0.2~\cite{Kim2012PRB, Allen2016APL, Niedermeier2017PRB} for doped $\mathrm{BaSnO_3}$. Our results suggest a relatively small $m^*$ at the conduction band bottom and a large non-parabolicity. For $N$ above \SI{1E20}{cm^{-3}}, our results are in good agreement with the value reported in Ref.~\cite{Allen2016APL}, which is obtained from Hall and infrared reflectivity measurements and is widely accepted for $\mathrm{BaSnO_3}$. The quick increase of $m^*$ with $N$ we find here is also in line with a recent theoretical study~\cite{Rowberg2020PRB} (green curve in Fig.~\ref{fig:4}\textbf{b} and \textbf{c}), where a conduction band inflection point is predicted to exist in at $\mathrm{BaSnO_3}$ large Fermi energies.

Given that the B-M shift and the bulk plasmon energy are both dependent on the band curvature, a comparison between the conduction band extracted above with the experimental bulk plasmon energy and B-M shift from EELS serves as a further verification. As shown in Fig.~\ref{fig:4}\textbf{d}, the relation between the B-M shift and the bulk plasma frequency can be described reasonably well with the non-parabolic conduction band model and parameters given above. We note that it should be also possible to extract the conduction band curvature by fitting the relation between the bulk plasmon energy and the B-M shift. 

Furthermore, combining the above analysis with core-loss excitations allows us to extract the doping activation percentage $p$, another important parameter for many material systems~\cite{Capano1998JEM, Tandon2019CM, Son2010NatureMaterials, Mairoser2010PRL, Held2020PRM}, which is defined by the ratio between carrier density and dopant concentrations ($p=NV/\mathrm{x}$, where $V$ is the unit cell volume). In practice, $p$ is usually evaluated from the ratio between Hall carrier density and nominal dopant concentration, which is however difficult to measure and not well understood at the microscopic level. In BLSO thin films studied earlier, $p$ was found to vary over a wide range, from $<10$\% to $\sim70$\%~\cite{Niedermeier2016APL,Kim2012PRB,Singh2019ACSAEM,Wang2019APLMaterials}.

In our study, the Hall carrier densities for the bulk single crystals with $\mathrm{x=}$ 0.01 and 0.05 correspond to $p=$ 80\% and 50\%, respectively, as shown in Fig.~\ref{fig:4}\textbf{e}. With the band curvature extracted above, the local carrier density can be calculated from either the B-M shift or the bulk plasma frequency. We use the former in our calculation, as the overlap between bulk and surface carrier plasmons renders the extraction of $\omega_{\rm p}$ less straightforward for low-doping regions. We note that the carrier densities calculated from the B-M shift and the bulk plasma frequency are in good agreement with each other, as shown in the bottom panel of Fig.~\ref{fig:2}\textbf{g}.

Combining the local carrier density with the La percentage shown in Fig.~\ref{fig:2}, we also show the activation percentage for the inhomogeneous nanocrystals in Fig.~\ref{fig:4}\textbf{e}, separately measured in 3.2 $\times$ 3.2 $\mathrm{nm^2}$ regions. We observe that the dopant densities in this nanocrystal vary over a broad range, between \SI{6E19} and \SI{2E21}{cm^{-3}},  while the carrier densities are centered around \SI{2E20}{cm^{-3}}. As a result, $p$ for the majority of the regions lies in the range between 25\% and 50\%. Overall, the dopant activation percentage decreases with doping level for both bulk and nanocrystals of BLSO. The primary source of electron donors are La dopants, as oxygen vacancies are not likely to form in $\mathrm{BaSnO_3}$~\cite{Wang2019PRM}. Thus, the percentage of La dopants that does not contribute to free carriers is rather high, at least 50\% for most of the regions. Possible reasons for the incomplete doping activation include charge 
compensation at surfaces and by defects~\cite{Scanlon2013PRB}.

\section{Discussion}
The present study demonstrates the usefulness of measuring low-energy excitations through EELS for understanding the electrical and optical properties at length scale of a few nanometers. We have shown that the conduction band curvature can be quantitatively modeled when the energy of both the bulk carrier plasmon and the band gap are known, thus allowing quantification of $N$ and $m^*$. The good agreement obtained between EELS analysis and other experimental methods (Hall effect, infrared reflectivity) suggest that EELS can reliably measure doping-induced changes in the bulk plasma frequency and optical gap and, when combined with transport measurements, it allows us to quantify local carrier densities. Changes in conduction band occupancy should, in principle, also be reflected in near-edge fine structures of core-loss EELS. However, it is impractical to measure such small changes with high accuracy by core-loss EELS due to the core-hole lifetime broadening and the small energy-loss cross-section at high energies. In comparison, low-energy excitations are more sensitive to electronic states near the Fermi level. Moreover, combining low-loss and core-loss EELS even allows for the doping activation percentage to be quantified, which otherwise can only be acquired from multiple measurements averaged over large areas. By exploiting the high spatial resolution STEM imaging, it should be possible to study local changes in electronic structure down to the individual-dopant level.

Our results also suggest that delocalization of inelastic scattering is not as large as previously perceived, especially for the bulk losses. It is well-known that higher-energy core-loss EELS provides information down to the sub-{\aa}ngstrom scale~\cite{Batson1993Nature,Muller2009NatureMaterials,Botton2010UM}, while lower-energy excitations are often considered to suffer from a poorer spatial localization~\cite{Egerton2017UM}. Here, we show that both free carrier plasmons and band gap transitions contain information that is sufficiently well localized as to yield useful insight into local optical properties (e.g., high/low-density plasmon trapping, see below), from which knowledge on the electronic structure (e.g., the effective mass) can be retrieved with a resolution of a few nanometers in our experiment, ultimately limited by the effective Bohr radius of the carrier electrons.

Remarkably, our analysis has revealed the existence of a previously unobserved plasmonic mode that is highly confined between regions with different doping level. This finding supports the possibility of engineering doped structures at the few-nm level for guiding electromagnetic waves along an arbitrary path defined by the doping profile, complementary to geometrical control of metallic nanostructures~\cite{Rossouw2013PRL, Raza2016NatureComm}. While this approach can be followed over a wide range of length scales depending on the light wavelength and degree of confinement (e.g., in metamaterials operating from microwaves to the visible), the present analysis relying on BLSO demonstrates a specific example operating in the mid-infrared regime with a high degree of spatial confinement by 2-3 orders of magnitude relative the light wavelength.

In brief, our work unlocks the full potential of EELS in providing unique microscopic insights on the structure-property relationship, here demonstrated by providing details on the electronic band structure including the energy-dependence of the electron mass as well as revealing a novel type of interface plasmons.

\section{Materials and Methods}
\subsection{STEM-EELS}
STEM-EELS data were acquired using a Nion Ultra STEM 100 equipped with an electron monochromator. In our EELS experiments, energy resolution and beam current were balanced for each energy range of interest. For phonons and plasmons in the infrared, a spectrometer dispersion of 1.2~meV/pixel was used to provide the highest possible energy resolution. For band-gap measurements, the spectrometer dispersion was set to 30~meV/pixel to improve the signal-to-noise ratio. This gave a full width at half maximum of the zero-loss peak (ZLP) of 33~meV. For core-loss EELS in the 400-900~eV region, we set the spectrometer dispersion to 300~meV/pixel and did not introduce the monochromator in order to maximize the beam current and thus, the signal-to-noise ratio. The EELS detector dwell time was set such that the ZLP is just below saturation.

\subsection{BLSO crystal growth and Hall measurement}
The $\mathrm{Ba_{1-x}La_xSnO_3}$ nanocrystals were synthesized by sol-gel method described in~\cite{Yang2022Small}. The bulk $\mathrm{Ba_{1-x}La_xSnO_3}$ crystals were grown by a laser floating zone technique. High-purity powder of $\mathrm{La_2O_3}$, $\mathrm{BaCO_3}$, and $\mathrm{SnO_2}$ were mixed in molar ratio x/2:1-x:1.5. $\mathrm{La_2O_3}$ was baked at 900$^{\circ}$C overnight before weighing. The mixed powder was calcinated at 1000$^{\circ}$C for 10 hours. The product was reground and sintered at 1400$^{\circ}$C for 20 hours with one intermediate grinding. The product was reground, filled into a rubber tube, and pressed under 8000 PSI hydrostatic pressure. The rod was sintered at 1400$^{\circ}$C for 10 hours. The crystals were grown at a rate of 50 mm/hr under 8 bar $\mathrm{O_2}$ pressure. The as-grown crystals were annealed at 1400$^{\circ}$C for 20 hours in oxygen flow before measurements.

For Hall measurements, the crystal was oriented, cut, and polished into a (100) plate. The electrodes were made by gold sputtering. The Hall measurements were performed in a Quantum Design PPMS-9 by sweeping the magnetic field at room temperature.

\subsection{EELS simulations}
We use a finite-element method (as implemented in COMSOL Multiphysics) to calculate the EELS probability for BLSO flakes, with the dielectric function made position-dependent through the spatial distribution of the doping density, as specified in the Supplementary Information.

\section*{Acknowledgement}
H.Y. and P.E.B. acknowledge the financial support of
the U.S. Department of Energy, Office of Science,
Basic Energy Sciences under Award No.~DESC0005132. A.K. acknowledges the ESF under the project CZ.02.2.69/0.0/0.0/20\_079. X.X. and S.W.C. were supported by the center for Quantum Materials Synthesis (cQMS), funded by the Gordon and Betty Moore Foundation’s EPiQS initiative through grant GBMF10104, and by Rutgers University. F.J.G.A. acknowledges support by the European Research Council (Advanced Grant 789104-eNANO), the Spanish MICINN (PID2020-112625GB-I00 and Severo Ochoa CEX2019-000910-S), and Fundaci\'os Cellex and Mir-Puig.

\bibliography{Dopant_carrier_refs}
\bibliographystyle{ieeetr}

\end{document}